\documentclass[twocolumn,trackchanges]{aastex7}
\usepackage{fix-cm}
\usepackage{multirow}
\usepackage{graphicx}	
\usepackage{amsmath}
\usepackage{makecell}
\usepackage{hyperref}
\usepackage{enumitem}
\usepackage{fontawesome5}
\usepackage{mathrsfs}
\setlist[itemize]{noitemsep, topsep=4pt}

\defcitealias{Minor2021}{M21}
\defcitealias{Minor2025}{M25}
\defcitealias{Nightingale2024}{N24}
\defcitealias{Enzi2024}{E25}
\defcitealias{Despali2024}{D24}
\defcitealias{Ballard2024}{B24}
\defcitealias{Vegetti2010}{V10}
\defcitealias{Tajalli2025}{T25}
\defcitealias{He2024}{H24}

\newcommand{\citeMA}{\citetalias{Minor2021}}
\newcommand{\citeMB}{\citetalias{Minor2025}}
\newcommand{\citeN}{\citetalias{Nightingale2024}}
\newcommand{\citeE}{\citetalias{Enzi2024}}
\newcommand{\citeD}{\citetalias{Despali2024}}
\newcommand{\citeB}{\citetalias{Ballard2024}}
\newcommand{\citeV}{\citetalias{Vegetti2010}}
\newcommand{\citeT}{\citetalias{Tajalli2025}}
\newcommand{\citeH}{\citetalias{He2024}}

\newcommand\fnnls{\texttt{fnnls}}

\begin{document}

\title{Not so dark, not so dense: an alternative explanation for the lensing subhalo in SDSSJ0946+1006}

\author[orcid=0000-0003-3672-9365,sname='He']{Qiuhan He}
\affiliation{Institute for Computational Cosmology, Department of Physics, Durham University, South Road, Durham DH1 3LE, UK}
\email{qiuhan.he@durham.ac.uk}

\author[0000-0002-0086-0524]{Andrew Robertson}
\affiliation{Carnegie Observatories, 813 Santa Barbara Street, Pasadena, CA 91101, USA}
\email{arobertson@carnegiescience.edu}

\newcommand{\ar}[1]{{\textcolor{orange}{\sf{[AR: #1]}} }}

\author[0000-0002-8987-7401]{James W. Nightingale}
\affiliation{School of Mathematics, Statistics and Physics, Newcastle University, Newcastle upon Tyne, NE1 7RU, UK}
\affiliation{Institute for Computational Cosmology, Department of Physics, Durham University, South Road, Durham DH1 3LE, UK}
\affiliation{Centre for Extragalactic Astronomy, Department of Physics, Durham University, South Road, Durham DH1 3LE, UK}
\email{James.Nightingale@newcastle.ac.uk}

\author[0000-0002-4465-1564]{Aristeidis Amvrosiadis}
\affiliation{Institute for Computational Cosmology, Department of Physics, Durham University, South Road, Durham DH1 3LE, UK}
\email{aristeidis.amvrosiadis@durham.ac.uk}

\author[0000-0002-5954-7903]{Shaun Cole}
\affiliation{Institute for Computational Cosmology, Department of Physics, Durham University, South Road, Durham DH1 3LE, UK}
\email{shaun.cole@durham.ac.uk}

\author[0000-0002-2338-716X]{Carlos S. Frenk}
\affiliation{Institute for Computational Cosmology, Department of Physics, Durham University, South Road, Durham DH1 3LE, UK}
\email{c.s.frenk@durham.ac.uk}

\author[0009-0007-0679-818X]{Samuel C. Lange}
\affiliation{Institute for Computational Cosmology, Department of Physics, Durham University, South Road, Durham DH1 3LE, UK}
\email{samuel.c.lange@durham.ac.uk}

\author[0009-0004-0904-7400]{Shubo Li}
\affiliation{School of Physics and Astronomy, Beijing Normal University, Beijing 100875, China}
\affiliation{National Astronomical Observatories, Chinese Academy of Sciences, 20A Datun Road, Chaoyang District, Beijing 100101, China}
\affiliation{School of Astronomy and Space Science, University of Chinese Academy of Sciences, Beijing 100049, China}
\email{lisb@nao.cas.cn}

\author[0000-0003-3899-0612]{Ran Li}
\affiliation{School of Physics and Astronomy, Beijing Normal University, Beijing 100875, China}
\email{liran@bnu.edu.cn}

\author[0000-0003-4988-9296]{Xiaoyue Cao}
\affiliation{School of Astronomy and Space Science, University of Chinese Academy of Sciences, Beijing 100049, China}
\affiliation{National Astronomical Observatories, Chinese Academy of Sciences, 20A Datun Road, Chaoyang District, Beijing 100101, China}
\affiliation{School of Physics and Astronomy, Beijing Normal University, Beijing 100875, China}
\email{xycao@nao.cas.cn}

\author{Leo W.H. Fung}
\affiliation{Institute for Computational Cosmology, Department of Physics, Durham University, South Road, Durham DH1 3LE, UK}
\email{}

\author{Xianghao Ma}
\affiliation{School of Physics and Astronomy, Beijing Normal University, Beijing 100875, China}
\affiliation{National Astronomical Observatories, Chinese Academy of Sciences, 20A Datun Road, Chaoyang District, Beijing 100101, China}
\affiliation{School of Astronomy and Space Science, University of Chinese Academy of Sciences, Beijing 100049, China}
\email{maxh@nao.cas.cn}

\author[0000-0002-6085-3780]{Richard Massey}
\affiliation{Institute for Computational Cosmology, Department of Physics, Durham University, South Road, Durham DH1 3LE, UK}
\email{r.j.massey@durham.ac.uk}

\author{Kaihao Wang}
\affiliation{School of Physics and Astronomy, Beijing Normal University, Beijing 100875, China}
\affiliation{National Astronomical Observatories, Chinese Academy of Sciences, 20A Datun Road, Chaoyang District, Beijing 100101, China}
\affiliation{School of Astronomy and Space Science, University of Chinese Academy of Sciences, Beijing 100049, China}
\email{wangkh@nao.cas.cn}

\author[0000-0003-4986-5091]{Maximilian von Wietersheim-Kramsta}
\affiliation{Institute for Computational Cosmology, Department of Physics, Durham University, South Road, Durham DH1 3LE, UK}
\affiliation{Centre for Extragalactic Astronomy, Department of Physics, Durham University, South Road, Durham DH1 3LE, UK}
\email{maximilian.von-wietersheim-kramsta@durham.ac.uk}

\correspondingauthor{Qiuhan He}
\email{qiuhan.he@durham.ac.uk}


\begin{abstract}

Previous studies of the strong lens system SDSSJ0946+1006 have reported a dark matter subhalo with an unusually high central density, potentially challenging the standard cold dark matter (CDM) paradigm. However, these analyses assumed the subhalo to be completely dark, neglecting the possibility that it may host a faint galaxy. In this work, we revisit the lensing analysis of SDSSJ0946+1006, explicitly modelling the subhalo as a luminous satellite. Incorporating light from the perturber broadens the range of allowed subhalo properties, revealing solutions with significantly lower central densities that are consistent with CDM expectations. The inferred luminosity of the satellite also aligns with predictions from hydrodynamical simulations. While high-concentration subhaloes remain allowed, they are no longer statistically preferred. The luminous subhalo model yields a better fit to the data, while also offering a more plausible explanation that is in line with theoretical expectations. We validate our methodology using mock data, demonstrating that neglecting subhalo light can lead to inferred mass distributions that are artificially compact.

\end{abstract}

\keywords{\uat{Dark matter}{353}; \uat{Strong gravitational lensing}{1643}; \uat{Galaxy dark matter halos}{1880}; \uat{Dwarf galaxies}{416}}


\section{Introduction} 
\setcounter{footnote}{0}

Cold dark matter (CDM) is the current standard model for dark matter (DM). One of its most fundamental predictions is the existence of a large population of dark matter haloes, ranging from massive galaxy clusters down to approximately the mass of the Earth \citep{Frenk2012, Wang2020, 2024MNRAS.528.7300Z}. Large galaxy redshift surveys \citep{Cole2005, Eisenstein2011, DESI_EDR} have revealed excellent agreement between the large-scale structure of our Universe and the predictions of CDM. However, on sub-galactic scales, $\lesssim10^{10}\,\mathrm{M_\odot}$, the accuracy of the theory remains largely untested, because at these mass scales galaxy formation is inefficient, such that haloes with these masses should host no, or very faint, galaxies \citep{Benitez-Llambay2020}. Though challenging, probing dark matter haloes in this low mass range would present a stringent test of the CDM theory, expanding our understanding of the nature of DM. 

Strong gravitational lensing is a powerful tool for probing invisible DM haloes at extragalactic distances. As predicted by general relativity, this phenomenon occurs when a massive object—such as a galaxy or cluster—lies along the line of sight to a more distant background source, bending and magnifying its light into multiple images, arcs, or rings. Small-scale DM haloes, either as \textit{subhaloes} within the main lens or \textit{field haloes} along the line of sight, can perturb these lensed features \citep{Vegetti2009b, Vegetti2010, Gilman2019, Gilman2020a, He2022}. Such perturbations, detectable through detailed lens modelling, offer a unique window into the DM distribution on sub-galactic scales \citep{Vegetti2014_SENS, Vegetti2018, Li2016, Li2017, Despali2017, Despali2018, Amorisco2022, He2022_SENS}. To date, several individual sub-galactic DM haloes have been detected through strong lensing: two in image data, SDSSJ0946+1006 \citep{Vegetti2010} and JVAS B1938+666 \citep{2012Natur.481..341V}, one in radio data, SDP-81 \citep{2016ApJ...823...37H} and an additional candidate through James Webb Space Telescope (JWST), SPT-2147 \citep{Lange2025}.

Among those detections, the subhalo found in SDSSJ0946+1006 has attracted the most attention due to its unique properties. Using both parametric and pixelized lens mass reconstruction techniques \citep{Koopmans2005, Vegetti2009a, Cao2025}, \citet{Vegetti2010} (hereafter \citeV{}) first detected a subhalo with a mass of $\sim3.5\times10^9\,\mathrm{M_\odot}$ \citep[assuming a pseudo-Jaffe mass profile][]{Jaffe1983, Munoz2001} in the HST F814W (I-band) image \citep{Bolton2006}. Follow-up analysis by \citet{Minor2021} (hereafter \citeMA{}) independently confirmed the detection and revealed that the subhalo exhibits an extremely compact structure: under the assumption of a truncated Navarro-Frenk-White (tNFW) profile \citep{NFW1996, Baltz2009}, a concentration exceeding 500 (20 to 50 times larger than expected!) was measured, with a projected mass of $\sim3\times10^9\,\mathrm{M_\odot}$ enclosed within a radius of 1~kpc. Even after adopting a more complex mass model for the main lens galaxy (incorporating multipole perturbations \citep{Powell2022, Stacey2024, Amvrosiadis2024}) the inferred concentration remains as high as 70 — still deviating from the concentration expected for a pure dark halo in CDM universe by more than $5\mathrm{\sigma}$. Subsequent studies have consistently corroborated this result, using a variety of independent methods and modelling approaches \citep[hereafter \citeN{}, \citeB{}, \citeD{}, \citeMB{}, \citeE{} and \citeT{} respectively]{Nightingale2024, Ballard2024, Despali2024, Minor2025, Enzi2024, Tajalli2025}.

The super-concentrated dark perturber detected in SDSSJ0946+1006 poses a significant challenge to CDM and has sparked considerable discussion. One possible mechanism to produce high central densities in a CDM halo is the condensation of baryons at the halo centre, i.e. gas cooling with the subsequent formation of a galaxy. The mass of the stars and gas increases the total matter density (which is directly probed by lensing), and can also increase the central DM density through a process referred to as adiabatic contraction \citep{1986ApJ...301...27B, 2004ApJ...616...16G}. However, previous analyses have constrained the luminosity of such a hypothetical galaxy to be less than $\sim10^8\,\mathrm{L_\odot}$ (\citeMA{}),\footnote{Note that \citeV{} claimed a tighter upper limit of $5 \times 10^6\,\mathrm{L_\odot}$, from the absence of positive residuals (data minus model) at the perturber location in their best-fitting lens model. This is likely an under-estimate though, because the model for the gravitationally lensed source light will try to adapt to reduce any residuals.} which would be insufficient to significantly alter the DM distribution, and therefore would not explain the high inferred concentration. 

A more exotic explanation is that this detection points towards DM being something other than CDM. In particular, in a self-interacting dark matter (SIDM) scenario, subhaloes can undergo a process known as core collapse \citep[e.g.][]{2002ApJ...568..475B}, leading to very high central DM densities. This possibility has been explored by several studies (\citeMA{}, \citet{2023ApJ...958L..39N}, \citeD{}, \citeMB{}, \citeE{} and \citeT{}). However, \citet{Li2025} have recently shown that even for a collapsed SIDM subhalo, achieving an enclosed mass of $3\times10^9\,\mathrm{M_\odot}$ within 1~kpc would require an original halo mass exceeding $5\times10^{10}\,\mathrm{M_\odot}$. Such a massive halo would be expected to host a galaxy bright enough to be detected in the imaging data, yet no such luminous counterpart has been found, challenging the SIDM interpretation.

It is also possible that the perturber is not associated with the lens galaxy itself but is instead located along the line of sight. However, \citet{He2022_SENS} pointed out that the lensing efficiency is maximized when the perturber lies close to the plane of the lens galaxy; if it were located in front of or behind the main lens, an even larger mass would be required to produce the same lensing signal, making the presence of a detectable galaxy even more likely. Furthermore, recent results by \citeE{} and \citeT{} also suggest that the perturber is likely to be a subhalo of the lens galaxy.

Given the significant challenge posed by this perturber, it is worth considering whether there are mechanisms that could lead to its mass distribution being incorrectly inferred. Here, we consider whether light associated with the lensing perturber could (if not included in the model) lead to the perturber's density being overestimated. In prior work (\citeV{}, \citeMA{}, \citeN{}, \citeB{}, \citeD{}, \citeMB{}, \citeE{} and \citeT{}), the preference for a subhalo is driven by the fact that it removes a localised mismatch between the data and perturber-free models (e.g. see Figure A.1 of \citeD{}).

Most prior work has fit a mass distribution for the perturber assuming that the residuals are caused solely by its gravitational lensing effect, and have then subsequently shown that the resulting model fits the data well, leaving little room for the perturber to have its own light associated with it. In reality, if the perturber hosts a faint galaxy then the inferred source surface brightness distribution, and lens mass distribution will adjust themselves to try and best fit the data, potentially ``absorbing'' the signal of any perturber light. To accurately investigate the possibility of perturber light, one needs to simultaneously model both the lensing effects of the perturber, as well as any possible light contribution. 
Only \citeMA{} have considered this possibility in their analysis; but, due to limitations in model flexibility -- in particular, the pre-subtraction of the foreground galaxy’s light -- they did not explore this possibility in detail.


In this work, we reanalyze the SDSSJ0946+1006 lensing system. As in previous studies, we focus primarily on the HST F814W observation, since the previous detections of the subhalo have been primarily based on this dataset. We employ our state-of-the-art, open-source lens modelling software, \textsc{PyAutoLens} \href{https://github.com/Jammy2211/PyAutoLens}{\faGithub} \citep{Nightingale2018}, to perform a comprehensive analysis of this system. In modelling the effects of the subhalo on the lensed images, we not only adopt the previous approach (where only the lensing perturbation from the subhalo's mass is considered) but also account for the possibility of light from a galaxy hosted at the centre of the subhalo. We demonstrate that once this light emission is included, the inferred mass distribution of the subhalo becomes fully consistent with the predictions of the CDM model, eliminating the need for an extremely compact mass profile to explain the observed signal. Taking into account the potential contribution from the light, we detect a subhalo with a mass of $\log_{10}\left(m_{200}/\mathrm{M}_\odot\right)=9.5^{+0.7}_{-1.1}$ (hereafter, unless specified otherwise, errors are quoted at the 95\% confidence level), a luminosity of $\log_{10}\left(L/\mathrm{L}_{\odot}\right)=8.4^{+0.1}_{-0.2}$, a reasonale amount of light for a subhalo of this mass. The enclosed subhalo mass within 1~kpc is reduced to be $\log_{10}(M_\mathrm{1kpc}/\mathrm{M_\odot})=8.9_{-0.8}^{+1.1}$. Compared to the high-density mass-only subhalo solution, our model with an additional light component provides a significantly better fit to the data, with an increase in Bayesian evidence of 16, corresponding to a $>5\mathrm{\sigma}$ preference for including subhalo light. 


The structure of this paper is as follows: In Sec.~\ref{sec:data}, we describe the observational data used. Sec.~\ref{sec:modelling_pipe} presents our lens modelling procedure, including the methodology and the models employed. In Sec.~\ref{sec:results}, we present the main results of our analysis. To validate the robustness of our modelling, we also test our modelling pipeline using simulated datasets in Sec.~\ref{sec:mocks}. We then discuss the implications of our findings in Sec.~\ref{sec:discussion}. Finally, we summarize our main conclusions in Sec.~\ref{sec:conclusions}. Throughout the work, we adopt the Planck $\Lambda$CDM cosmology \citep{Planck2016}. The angular diameter distance to the lens galaxy is 761.3~Mpc, meaning that 1 arcsec corresponds to a distance of 3.69 kpc in the lens plane. The luminosity distance to the lens plane is 1136.8~Mpc.

\section{Data}\label{sec:data}
In this work, we analyze the F814W-band HST image of SDSSJ0946+1006 (HST Proposal ID: 10886), processed following the same procedure as used in \citet{Etherington2022} and \citeN{}. The image has a pixel scale of 0.05\arcsec and the model point spread function (PSF) is generated using the \textsc{Tiny Tim} software \citep{Krist1993}. This system is known to have three lensed sources at different redshifts \citep{Gavazzi2007, Collett2020}. In this work, we only model the innermost lensed arcs (with an Einstein radius of $\sim1.4\arcsec$, $z_\mathrm{lens}=0.222$ and $z_\mathrm{source}=0.609$), where the subhalo falls. Accordingly, we apply a circular mask with a radius of 2.2\arcsec and manually mask out the region containing emission from the second lensed source, as done in \citeN{}. 

\section{Lens Modelling Pipeline}\label{sec:modelling_pipe}

\begin{table}
\renewcommand{\arraystretch}{1.0}
\setlength{\tabcolsep}{1.5pt} 
\centering
\small
\begin{tabular}{ccccc}
\hline
\textbf{Phase} & \textbf{Fit} & \textbf{Component} & \textbf{Model} & \textbf{Prior} \\ \hline
\multirow{3}{*}{\shortstack{Source\\Parametric}} 
& \multirow{3}{*}{\textbf{SP}}  
& Lens mass & $\text{SIE} + \text{Shear}$ & - \\ \cline{3-5} 
& & Lens light & MGE & - \\ \cline{3-5} 
& & Source light & MGE & - \\ \hline
\multirow{6}{*}{\shortstack{Source\\Pixelized}} 
& \multirow{3}{*}{\textbf{Pix1}} 
& Lens mass & $\text{SIE} + \text{Shear}$ & \textbf{SP} \\ \cline{3-5} 
& & Lens light & MGE & \textbf{SP} \\ \cline{3-5} 
& & Source light & Voronoi & - \\ \cline{2-5} 
& \multirow{3}{*}{\textbf{Pix2}} 
& Lens mass & $\text{SIE} + \text{Shear}$ & \textbf{Pix1} \\ \cline{3-5} 
& & Lens light & MGE & \textbf{SP} \\ \cline{3-5} 
& & Source light & Voronoi & - \\ \hline
\multirow{3}{*}{Light} 
& \multirow{3}{*}{\textbf{L}}  
& Lens mass & $\text{SIE} + \text{Shear}$ & \textbf{Pix1} \\ \cline{3-5} 
& & Lens light & MGE & \textbf{SP} \\ \cline{3-5} 
& & Source light & Voronoi & \textbf{Pix2} \\ \hline
\multirow{3}{*}{Mass} 
& \multirow{3}{*}{\textbf{M}}  
& Lens mass & $\text{EPL}+\text{Shear}$ & \textbf{Pix1} \\ \cline{3-5} 
& & Lens light & MGE & \textbf{L} \\ \cline{3-5} 
& & Source light & Voronoi & \textbf{Pix2} \\ \hline
\multirow{3}{*}{\shortstack{Source\\Tune}} 
& \multirow{3}{*}{\textbf{ST}}  
& Lens mass & $\text{EPL} + \text{Shear}$ & \textbf{M} \\ \cline{3-5} 
& & Lens light & MGE & \textbf{L} \\ \cline{3-5} 
& & Source light & Voronoi & - \\ \hline
\multirow{12}{*}{Subhalo} 
& \multirow{3}{*}{\textbf{\shortstack{No\\Subhalo}}} 
& Lens mass & $\text{EPL} + \text{Shear}$ & \textbf{M} \\ \cline{3-5} 
& & Lens light & MGE & \textbf{L} \\ \cline{3-5} 
& & Source light & Voronoi & \textbf{ST} \\ \cline{2-5} 
& \multirow{4}{*}{\textbf{\textbf{\shortstack{Dark\\Subhalo}}}} 
& Lens mass & $\text{EPL} + \text{Shear}$ & \textbf{M} \\ \cline{3-5} 
& & Lens light & MGE & \textbf{L} \\ \cline{3-5} 
& & Source light & Voronoi & \textbf{ST} \\ \cline{3-5} 
& & Subhalo mass & NFW & - \\ \cline{2-5} 
& \multirow{5}{*}{\textbf{\textbf{\shortstack{Luminous\\Subhalo}}}} 
& Lens mass & $\text{EPL} + \text{Shear}$ & \textbf{M} \\ \cline{3-5} 
& & Lens light & MGE & \textbf{L} \\ \cline{3-5} 
& & Source light & Voronoi & \textbf{ST} \\ \cline{3-5} 
& & Subhalo mass & NFW & - \\ \cline{3-5} 
& & Subhalo light & Sérsic & - \\ \hline
\end{tabular}
\caption{The Source, Light and Mass (SLaM) pipelines used in this analysis, built using \textsc{PyAutoLens}.}
\label{tab: lensing_pipeline}
\end{table}

We perform our strong lensing analyses using \textsc{PyAutoLens} \citep{Nightingale2015, Nightingale2018, Nightingale2021}. To reduce difficulties associated with initializing a complex model, we adopt the SLaM (Source, Light, and Mass) pipeline provided by \textsc{PyAutoLens}, which decomposes the modelling process into a sequence of phases. This approach starts from simple models, which then gradually increases in complexity. The SLaM pipeline has been extensively applied in previous studies and has demonstrated robust parameter inference for complex lensing configurations \citep{Cao2021, Etherington2022, Nightingale2024, He2023, He2024, Wang2025, Nightingale2025}. For our purposes, we modify and extend the standard SLaM pipeline to better suit our analysis, which is focused on constraining the subhalo's properties. To sample from the complex posterior distributions that arise in lensing, we use the nested samplers: \textsc{Dynesty} \citep{Speagle2020, Sergey2023} and \textsc{Nautilus} \citep{Lange2023}. A summary of the modelling pipeline used throughout this work is presented in Table~\ref{tab: lensing_pipeline}. In the following, we provide a detailed description of each modelling phase and the corresponding lens models used.

\subsection{Source Parametric (SP) phase}
The \textbf{SP} phase aims to obtain a robust lens mass model (e.g. constraining the mass centre and Einstein radius) to prevent unphysical demagnified pixelised source reconstructions in later phases \citep{Maresca2021}.

Both the lens and source light are modelled using a multi-Gaussian expansion (MGE), where their surface brightness distributions are represented by a sum of two-dimensional Gaussian components \citep{Cappellari2002}. The MGE approach provides a flexible and analytically convenient way to describe the emission profiles of foreground and background galaxies, and its implementation in \textsc{PyAutoLens} is detailed in \citet[][hereafter \citeH{}]{He2024}.

In this initial modelling stage, the lens light is described using two MGE \emph{groups}, each comprising 30 individual Gaussian profiles. The Gaussians within each group have different scale radii, but share a common centre, position angle, and axis ratio. The scale radii, the standard deviations that set the spatial extent of each Gaussian component, are fixed and logarithmically-spaced between 0.01\arcsec and the mask radius of 2.2\arcsec.

Depending on the model complexity, different MGE groups may be assigned different centres, position angles and axis ratios; however at this early stage, we require the two groups to share the same centre. As described later in the \textbf{Light} phase, we will relax this constraint to allow for enhanced flexibility in the lens-light model. The source light is similarly modelled with one MGE group containing 30 Gaussian profiles, with scale radii logarithmically spaced between 0.01\arcsec and 0.1\arcsec. All the parameters except for the Gaussian intensities are sampled using a non-linear optimization algorithm, while the intensities are determined using a fast non-negative least squares (\fnnls) algorithm \citep{Bro1997}\footnote{The \fnnls~code used in this work is adapted from \url{https://github.com/jvendrow/fnnls}.}.

The lens mass is modelled as a singular isothermal ellipsoid (SIE), a special case of an elliptical power-law (EPL) mass profile with a convergence as \citep{Tessore2015}:
\begin{equation}
    \kappa(R) =
    \frac{3-\gamma}{1+q}\left(\frac{R_{\rm E}}{R}\right)^{\gamma-1},
 \label{equ:kappa_pl}
\end{equation}
where $R_{\rm E}$ is the Einstein radius, $q$ is the axis ratio and $\gamma$ is the slope of the density profile. The SIE corresponds to the case where $\gamma=2$. Ellipticity is introduced via the elliptical radius, $R$, as: 
\begin{equation}
    \begin{split}
        & R(x, y) = \sqrt{x^{\prime2} + \left(\frac{y^\prime}{q}\right)^2} , \\
        & x^{\prime} = \cos{\theta}\cdot\left(x - x^{\rm c}\right) + \sin{\theta}\cdot\left(y - y^{\rm c}\right) , \\
        & y^{\prime} = \cos{\theta}\cdot\left(y - y^{\rm c}\right) -\sin{\theta}\cdot\left(x - x^{\rm c}\right).
    \end{split}\label{eq: ellp}
\end{equation}
Here, $\theta$, $\left(x^{\rm c},\ y^{\rm c}\right)$ are the profile's position angle and centre, respectively.

We also include an external shear component in the mass model, parametrized by two components, $\gamma^{\rm ext}_1$ and $\gamma^{\rm ext}_2$,
\begin{equation}\label{eq:shear}
    \gamma^{\rm ext} = \sqrt{\gamma_{\rm 1}^{\rm ext^{2}}+\gamma_{\rm 2}^{\rm ext^{2}}}, \, \,
     \tan{2\phi^{\rm ext}} = \frac{\gamma_{\rm 2}^{\rm ext}}{\gamma_{\rm 1}^{\rm ext}},
\end{equation}
where $\gamma^{\rm ext}$ is the shear strength and $\phi^{\rm ext}$ is its position angle.

\subsection{Source Pixelized (Pix) phase}
In this phase, we model the source galaxy using a more flexible, pixelized Voronoi mesh representation. This is performed in two stages. In the first stage, \textbf{Pix1}, the Voronoi mesh centres are initialized based on the magnification pattern of the mass model (so that smaller pixels are used in regions of the source plane with greater magnification). In the second stage, \textbf{Pix2}, the centres are adapted to the lensed source morphology by clustering pixels according to the brightness distribution of the lensed emission \citep{Nightingale2018}. This is achieved using a newly developed clustering algorithm (\texttt{HilbertMesh} in \textsc{PyAutoLens}), which uses a Hilbert space-filling curve \citep{Hilbert1891} to optimize pixel placement. By assigning more pixels to brighter regions of the source, this scheme significantly improves the resolution with which we can reconstruct compact and cuspy sources \citep{Wang2025}. Once the lens mass model is fixed, a mapping is established between the pixelized source plane and the observed image data. The source intensities on the Voronoi mesh are interpolated using a natural-neighbour interpolation scheme \citep{Sibson1981} and regularized with a cross-like regularization structure. This approach addresses the noisy and stochastic likelihood systematics identified by \citet{Etherington2022} and \citeD{}, ensuring robust error estimates for the lens mass parameters. We refer readers to the Appendix of \citeH{} for a complete description of the natural-neighbour interpolated pixelized source reconstruction used here. 

Throughout this phase, we fix the lens mass and light profiles to the best-fit values obtained from the \textbf{SP} phase, except that the intensities of the MGE lens light model are re-optimized jointly with the source pixel fluxes using \fnnls.  

\subsection{Light (L) phase}
In this phase, we achieve a cleaner deblending of the lens and source light, enabled by the improved pixelized source reconstruction from the \textbf{SPix} phase. We refit a more complex MGE lens light model consisting of 6 MGE groups, each containing 10 Gaussian components. As in the \textbf{SP} phase, each group is allowed to have its own position angle and axis ratio, while the scale radii of the Gaussians are logarithmically spaced between 0.01\arcsec and the mask radius. Unlike in the \textbf{SP}, we now assume two distinct centres for the 6 MGE groups, with sets of 3 groups sharing a common centre. This allows for the modelling of potential centroid offsets in the lens light distribution, e.g. lopsideness \citep{Amvrosiadis2024}. Additionally, we introduce an extra `point-like' MGE group, composed of 10 Gaussians with scale radii values logarithmically spaced between 0.01\arcsec and 0.1\arcsec~to capture compact residuals at the galaxy centre. These could arise from slight mismatches between the model and true PSF, or from faint nuclear stellar emissions. 

Throughout this phase, the lens mass model and source parameters are fixed to the best-fit values obtained from \textbf{Pix2}. However, the MGE intensities and source pixel fluxes are still jointly optimized at each step using \fnnls. 

\subsection{Mass (M) phase}
The \textbf{Mass} phase is designed to fit a more complex macro lens mass distribution than the SIE model used previously. Depending on the scientific objectives, one can adopt different levels of complexity, such as modelling the macro lens mass with one or two EPLs, or introducing higher-order multipole terms to capture deviations from pure ellipticity \citep[e.g.][]{Riordan2023, Stacey2024}. Some previous studies have incorporated third- and fourth-order multipoles to model features such as boxiness or diskyness in the lens mass distribution, typically demonstrating an improved match between the best-fit model and the data. However, we note that previous work on SDSSJ0946+1006 has found that the inference of a super-concentrated subhalo remains even when adding multipoles to the main lens mass distribution (e.g. \citeMA{}, \citeMB{} and \citeT{}). Moreover, the amplitudes of these multipoles are typically constrained to very small values ($\lesssim 1\%$) when the second lensed source is also modeled (e.g., \citeMB{}, \citeE{}). Consequently, we adopt the simple approach of modelling the macro lens mass with a single EPL profile. During this phase, the MGE lens light model and source reconstruction parameters are fixed to the best-fit values from the \textbf{L} and \textbf{SPix} phases, except that their intensities are re-optimized through linear inversion for each mass model evaluated.

\subsection{Source Tune (ST) phase}
After obtaining optimal models for the lens mass and light distributions, we proceed to re-tune the regularization strength applied to the pixelized source reconstruction, prior to including perturbations from a subhalo. This step aims to maximize the capability of the pixelized source in capturing complex source features. As in the previous phase, the lens light and mass models are fixed to their best-fit values, except that the intensities of the lens light MGEs and source pixels are jointly re-optimised. 

\subsection{Subhalo phase}
In this phase, we focus on modelling the perturbation induced by a subhalo. We explore three different lens models:
\begin{itemize}
    \item \textbf{No Subhalo} --- We fit only an EPL profile to represent the macro lens mass distribution. This model serves as a baseline for comparison with the subsequent models that include a subhalo component. 
    \item \textbf{Dark Subhalo} --- In addition to the EPL profile, we introduce an NFW subhalo mass component to the lens model. The NFW profile is given by:
    \begin{equation}
        \rho(r) = \frac{\rho_s}{(r/R_s) (1 + r/R_s)^2}.
    \end{equation}
    where $\rho_s$ and $r_{\rm s}$ are the characteristic density and scale radius. For easier comparison with previous studies, we re-parameterize the profile in terms of its mass $m_{200}$ and concentration $c$. Here, $m_{200}$ denotes the mass enclosed within $r_{200}$, the radius within which the mean density is 200 times the critical density of the Universe. The concentration $c$ is defined as $r_{200} / r_{\rm s}$, with (at fixed $m_{200}$) higher concentrations implying higher central densities. Throughout this work, we assume that the perturber is at the same redshift as the main lens galaxy, which was shown to best explain the data by \citeE{} and \citeT{}.
    \item \textbf{Luminous Subhalo} --- This model adopts the same mass distribution as the dark subhalo model, a combination of an EPL profile for the main lens with an NFW subhalo. However, to account for potential stellar light associated with the subhalo, we additionally fit an elliptical S\'ersic light profile, with surface brightness profile:
    \begin{equation}
        I(r) = I_{\rm e}\cdot{\rm exp}\left[- b_n\left(\left(\frac{r}{r_{\rm e}}\right)^{1/n} - 1\right)\right],
    \end{equation}
    where $I_{\rm e}$ is the intensity at the effective radius $r_{\rm e}$, $n$ is the S\'ersic index and $b_n$ is a normalization factor determined by $n$ \citep{Graham2005}. The ellipticity is introduced to the profile following Eq.~(\ref{eq: ellp}). During the fitting, we require the subhalo mass and light components to have the same centre, as expected for a subhalo and its associated satellite galaxy. In total, this luminous subhalo model introduces five additional free parameters. 
\end{itemize}

Except for the differences in the subhalo models described above, all other model components and the prior distributions for the various parameters are kept identical across the three \emph{subhalo} fits. Specifically, the same MGE lens light and pixelized source models are used for each case. The priors used are summarized in Table~\ref{tab: priors} in Appendix. This phase is the final step of our analysis, and unless otherwise specified, all results presented in this paper are based on models fitted in this phase.

\section{Fitting results to the F814W image}\label{sec:results}

\begin{figure*}
    \includegraphics[width=2.0\columnwidth]{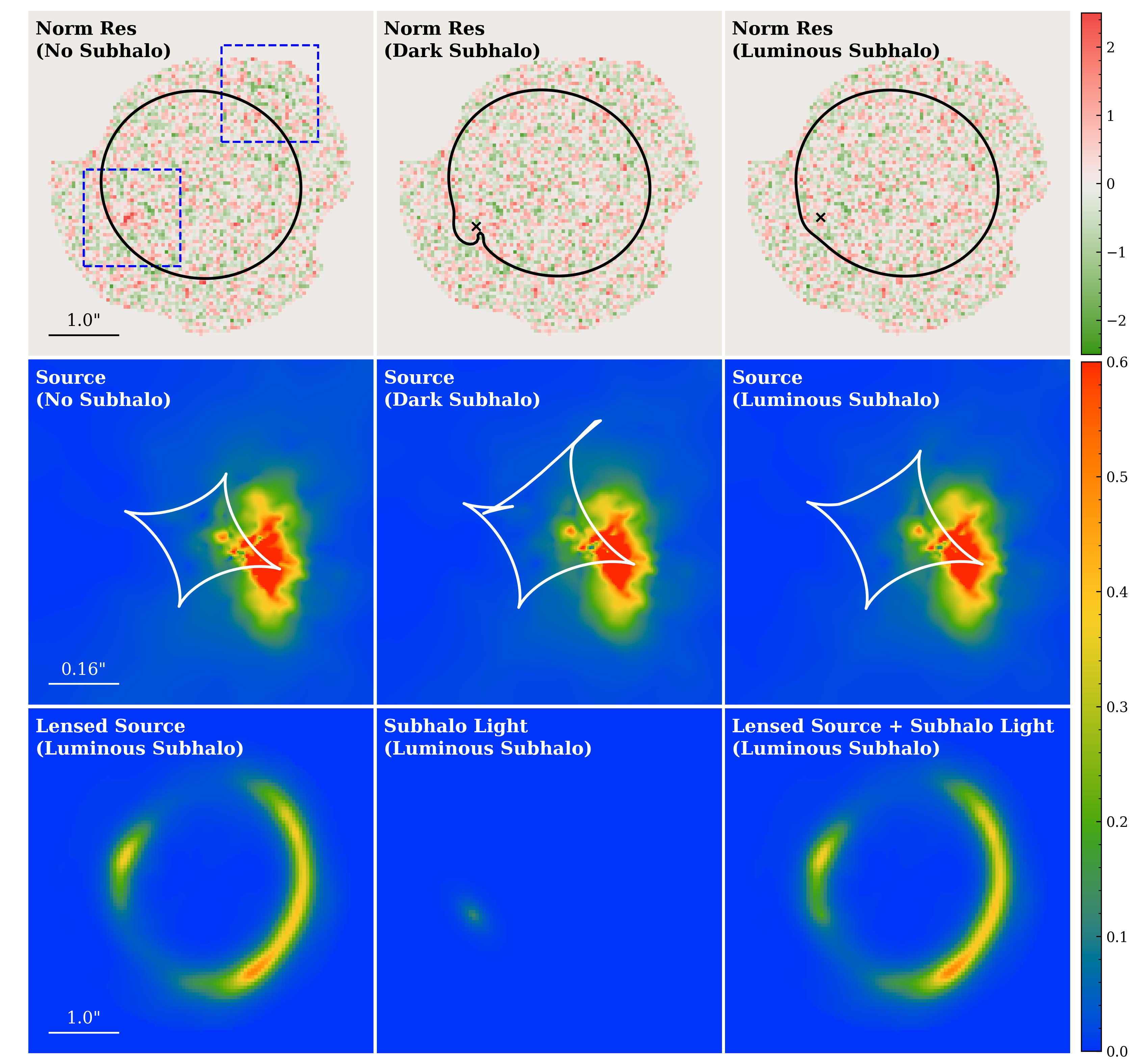}
    \caption{\textbf{Top:} The normalized residuals (i.e. data minus model, divided by noise) for three different best-fit lens models fit to the F814W image (left to right corresponds to the no-subhalo, dark subhalo and luminous subhalo cases, respectively. The dashed blue squares in the left panel highlight regions with large residuals. \textbf{Middle:} The best-fit source reconstructions of the three different subhalo models. The white lines mark the caustic lines on the source plane. \textbf{Bottom:} Comparison between the lensed source light and the subhalo light of the best-fit luminous subhalo model. The left is the lensed source emission, the middle is the subhalo light and the right shows the emission combining those two together. The images shown in this row have been convolved with the realistic PSF.}
    \label{fig: normalized_residuals}
\end{figure*}

In this section, we present the fitting results of the three different models in the \textbf{Subhalo} phase: the no-subhalo, dark subhalo and luminous subhalo.

\subsection{No-subhalo results}
We first show the macro-fitting results where no subhalo is included in the modelling. The top left panel of Fig.~\ref{fig: normalized_residuals} shows the best-fit normalized residuals ((data - model) / noises) of this no-subhalo model. As shown, the macro mass model (EPL + shear) provides a good fit across most of the image, with residuals generally at the noise level. It is noted that with the ``offset-centre'' MGE lens light model, we successfully fit the significant dipole residuals in the very central part of the lens galaxy reported in \citeH{}, \citeE{} and \citeMB{}. However, as pointed out by previous work, clear correlated residuals (highlighted by dashed blue squares in the panel) remain in the lower part of the left arc and the upper part of the right arc. As found previously, these residuals could plausibly be explained by the presence of a super-concentrated dark matter subhalo, which we now demonstrate.

\subsection{Dark subhalo results}
We next include an additional spherical NFW subhalo in the lens model, with the subhalo's location, mass, and concentration left as free parameters. The inferred mass for the NFW subhalo is $\log_{10}\left(m_{200}/{\rm M}_\odot\right)=9.8_{-0.3}^{+0.5}$ with a concentration of $\log_{10}\,c=2.5_{-0.5}^{+0.7}$, which is 6–14$\, \sigma$ higher than the median predicted by the mass–concentration relation in CDM \citep[e.g.][]{Ludlow2016}. 

\begin{figure}
    \includegraphics[width=1.0\columnwidth]{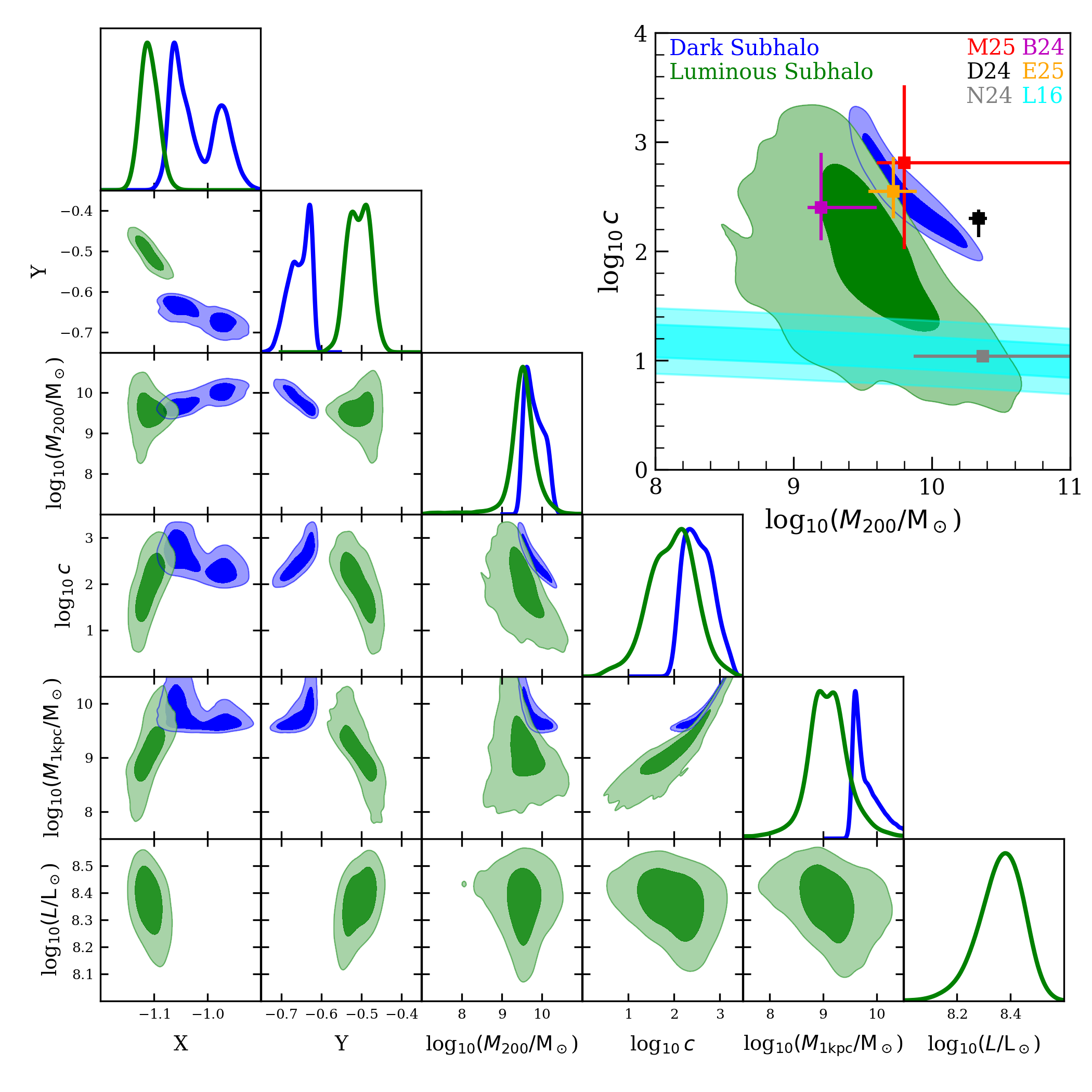}
    \caption{Projections of the posterior distribution for the subhalo parameters from both our dark subhalo (blue) and luminous subhalo (green) fits to the F814W image of SDSSJ0946+1006. The top-right panel shows the $M_{200} - c$ projection of the posterior, with previous measurements of the perturber's properties overlaid. Our dark subhalo model leads to an inferred mass and concentration for the subhalo in reasonable agreement with previous work (\citeD{}, \citeE{}, \citeB{}, \citeMB{}). The L16 band shows the expected concentration-mass relation in a CDM universe, taken from \citet{Ludlow2016}. In \citeN{}, the perturber was forced to lie on the L16 relation. Our luminous model decreases the concentration values required to explain the data.}
    \label{fig: subhalo_contours}
\end{figure}

The posterior distributions of the subhalo parameters are shown in Fig.~\ref{fig: subhalo_contours}, with 68\% and 95\% credible intervals shaded in dark and light blue, respectively. For the two key parameters of interest, $m_{200}$ and $c$, we compare our measurement with previous work in the upper-right panel of the figure. As shown, our constraints are generally consistent with previous findings, indicating a subhalo with an extremely high concentration (except for \citeN{} where $c$ is fixed to the median value expected in a CDM universe). Compared with the CDM expectation (represented by the shaded cyan region), it is clear that the lensing data favours a substantially more concentrated subhalo. Also, as done in some previous work, we compute the subhalo's projected mass within 1~kpc. We find this \emph{aperture mass} to be $\log_{10}\left(m_{1{\rm kpc}}/{\rm M}_\odot\right)=9.7_{-0.2}^{+0.9}$, consistent with the findings of \citeMB{}, \citeD{} and \citeE{}.

The middle panel of Fig.~\ref{fig: normalized_residuals} displays the normalized residuals of the best-fit dark subhalo model, which has a concentration of 740. The black cross marks the subhalo location, and the solid black line denotes the critical curve, which exhibits a distinct bump near the subhalo. As shown, the super-concentrated subhalo successfully fits the localized residuals in the lower (upper) region of the left (right) arc. Compared with no-subhalo model, the maximum log-likelihood improves by 70 and the log Bayesian evidence increases by 60, corresponding to a subhalo detection significance of over $10\, \sigma$.

\subsection{Luminous subhalo results}

In this subsection, we present results of the luminous subhalo model where we also take into account of the potential light contribution from the subhalo. In Fig.~\ref{fig: subhalo_contours}, we show the posterior distributions of the key parameters of the luminous subhalo model, plotted in deep and light green for the 68\% and 95\% credible intervals, respectively (the posterior distributions of all the parameters related to the subhalo's mass and light profiles can be found in Appendix in Fig.~\ref{fig: posteriors_subhalo_light}).

As shown by the posteriors, the new model measures the subhalo to have a mass of log$_{10}(m_{200}/{\rm M}_\odot)=9.5^{+0.7}_{-1.1}$ and a total luminosity of $\log_{10}(L/\mathrm{L}_\odot)=8.4_{-0.2}^{+0.1}$. It is noted that now \textbf{the median posterior concentration drops by around an order of magnitude and the concentration also becomes less tightly constrained, with $\log_{10}\,c=1.7_{-0.9}^{+1.2}$. This means that a much less concentrated mass profile is allowed for explaining the data once perturber-light is included in the model.} In the upper right panel of Fig.~\ref{fig: subhalo_contours}, we compare our new constraints on the subhalo's $m_{200}$ and $c$ with results from other models. We can see that inferred subhalo concentration is now fully compatible with the CDM expectation. In terms of the projected mass within 1~kpc, when perturber-light is taken into account, $M_{\rm 1kpc}$ is reduced to $\log_{10}(M_{\rm 1kpc}/{\rm M}_\odot)=8.9_{-0.8}^{+1.1}$, much smaller than that in the case of a \emph{dark} subhalo.

In terms of how well this luminous subhalo model fits the data, the right panel of Fig.~\ref{fig: normalized_residuals} shows the normalized residuals for the best-fit lunminous subhalo model, which has a concentration of 54. As in the previous model, the localized residuals are well fitted, but now the critical curve is considerably less perturbed. In terms of model comparison, \textbf{the model with both subhalo mass and light further improves the maximum log-likelihood by 23 compared with the model with a massive (but dark) subhalo, with a log Bayesian evidence increase of 16}, corresponding to a $>5 \, \sigma$ preference over the dark subhalo model.\footnote{In terms of detailed fitting statistics, the $\chi^2$ value of the best-fit luminous subhalo model is 23 lower than that of the best-fit dark subhalo model, indicating that the best-fit luminous subhalo model provides a better fit to the image. However, the regularization term of the best-fit luminous subhalo model is only 1 unit larger than that of the dark subhalo model, suggesting that the source smoothness in the two best-fit models is quite comparable.} Finally, in Table~\ref{tab: summary_fittings}, we summarize the model fitting results of the three different subhalo models we applied to the F814W image. 

\begin{table}
    \centering
    \renewcommand{\arraystretch}{1.2}
    \setlength{\tabcolsep}{2.0pt}
    \begin{tabular}{cccccc} 
    \hline
    Model & $\log_{10}m_{200}/{\rm M}_\odot$ & $\log_{10}c$ & $\log_{10}L_{\rm tot} / {\rm L}_\odot$ & $\mathcal{B}$ & $\Delta\mathcal{B}$ \\
    \hline
    \multirow{2}{*}{\shortstack{No\\Subhalo}} & \multirow{2}{*}{-} & \multirow{2}{*}{-} & \multirow{2}{*}{-} & \multirow{2}{*}{-1746} & \multirow{2}{*}{0} \\
    \\
    \hline
    \multirow{2}{*}{\shortstack{Dark\\Subhalo}} & \multirow{2}{*}{$9.8^{+0.5}_{-0.3}$} & \multirow{2}{*}{$2.5^{+0.7}_{-0.5}$} & \multirow{2}{*}{-} & \multirow{2}{*}{-1686} & \multirow{2}{*}{60} \\
    \\
    \hline
    \multirow{2}{*}{\shortstack{Luminous\\Subhalo}} & \multirow{2}{*}{$9.5^{+0.7}_{-1.1}$} & \multirow{2}{*}{$1.7^{+1.2}_{-0.9}$} & \multirow{2}{*}{$8.4^{+0.1}_{-0.2}$} & \multirow{2}{*}{-1670} & \multirow{2}{*}{76} \\
    \\
    \hline
    \end{tabular}
    \caption{A summary of the inferred properties of the subhalo from our different subhalo models. The first, second and third row shows the no-subhalo, dark subhalo, and luminous subhalo results. The log Bayesian evidence, $\mathcal{B}$, is listed for each model, with $\Delta\mathcal{B}$ the difference between $\mathcal{B}$ for the model in question and for the no-subhalo model. This means that $\Delta\mathcal{B}$ is the natural logarithm of the Bayes factor \citep{2008ConPh..49...71T}  when comparing each model to the no-subhalo case. The $\Delta\mathcal{B}$ values indicate that the dark subhalo model is preferred over the no-subhalo model, and the luminous subhalo model is preferred over the dark subhalo model.}\label{tab: summary_fittings} 
\end{table}

\section{Mock tests}\label{sec:mocks}

\begin{figure*}
    \includegraphics[width=2.0\columnwidth]{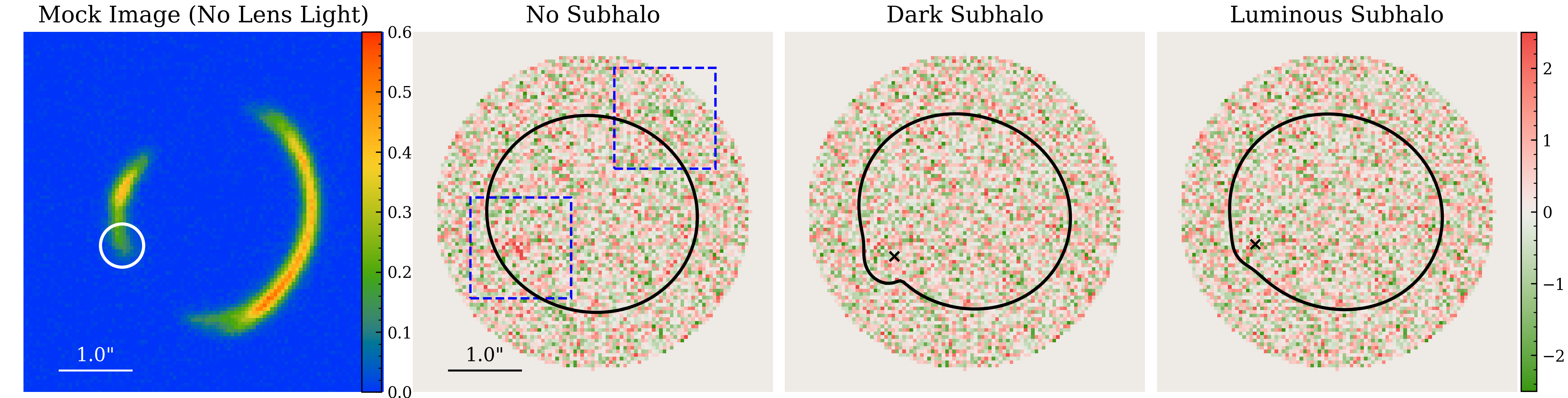}
    \caption{Normalized residuals for three different lens models fit to our mock data. \textbf{Left:} The normalized residual of best-fit no-subhalo model. The dashed blue squares highlight the localized resiudals corresponding to the subhalo. \textbf{Middle:} The normalized residuals of the dark subhalo model. \textbf{Right:} The normalized residuals of the best-fit luminous subhalo model.}
    \label{fig: mock_images}
\end{figure*}

Because the strong lens models applied in this work are complex, and the signals we are fitting to are subtle, it is important to check the reliability of our methodology against mock data. 
To do this, we simulate images representative of the F814W image of SDSSJ0946+1006 with similar noise levels, lens light distributions, macro lens mass distributions and source emission. For the (true) input subhalo, we assign it properties inspired by the best-fit parameters from the luminous subhalo model fit to the real observations, represented by an NFW halo plus an elliptical S\'ersic profile. Table~\ref{tab: mock_paras} summarizes the parameter of the input subhalo. The lens-light subtracted mock image is shown in the left panel of Fig.~\ref{fig: mock_images} (please note that the mock image used for tests do include lens light). The white circle marks the location of the input subhalo. As shown, the subhalo's light is very faint and does not appear as an obviously distinct component in the lens-light subtracted image. 

\begin{table}
    \renewcommand{\arraystretch}{1.2}
    \centering
    \begin{tabular}{ccc} 
    \hline
    \multirow{3}{*}{\begin{tabular}[c]{@{}c@{}}Mass Component: \\ spherical NFW \end{tabular}} & (x, y) & (-1.13\arcsec, -0.45\arcsec) \\
    \cline{2-3} & $m_{200}$ & $5.7\times10^9\,\mathrm{M}_\odot$ \\
    \cline{2-3} & $c$ & 32.0 \\
    \hline
    \multirow{5}{*}{\begin{tabular}[c]{@{}c@{}}Light Component: \\ elliptical S\'ersic \end{tabular}} & $I_\mathrm{e}$ & 0.05 e$^-$/pix \\
    \cline{2-3} & $R_\mathrm{e}$ & 0.16\arcsec \\
    \cline{2-3} & $n$ & 0.83 \\
    \cline{2-3} & $q$ & 0.61 \\
    \cline{2-3} & $\phi$ & $131^\circ$ \\
    \hline
    \end{tabular}
    \caption{Parameters of the input luminous subhalo in the mock image.}\label{tab: mock_paras} 
\end{table}

\begin{figure}
    \includegraphics[width=1.0\linewidth]{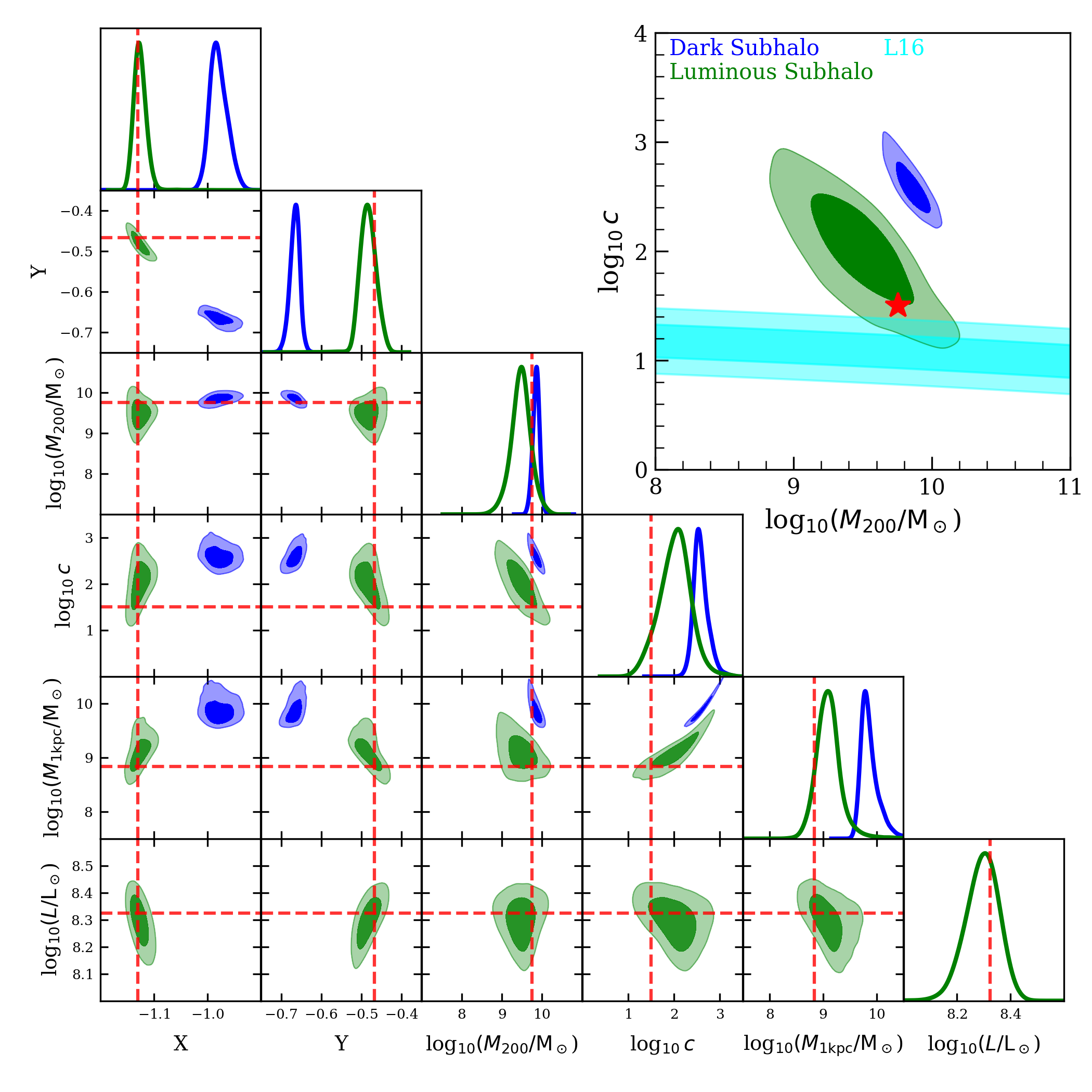}
    \caption{Posteriors of the subhalo model parameters of the mock test. Blue and green contours show the results of the dark and luminous subhalo model, respectively. The red dashed lines mark the true input values of the mock. Similarly, at the top right corner, we plot an zoom-in figure to show our constraints on the subhalo's mass and concentration in details.}
    \label{fig: posteriors_mock}
\end{figure}

We analyse the mock image using the same modelling pipeline as we applied to the real data. Again, three different models are fitted in the \textbf{Subhalo} phase: no-subhalo, dark subhalo, and luminous subhalo. Fig.~\ref{fig: posteriors_mock} shows the inferred posterior distribution for the subhalo's parameters for both dark subhalo (blue) and luminous subhalo (green) models (a full posterior of the luminous subhalo parameters can be found in Fig.~\ref{fig: posteriors_subhalo_light_mock} in Appendix). As shown, \textbf{the luminous subhalo model recovers the true input parameters of the subhalo, demonstrating that our method can simultaneously infer accurate distributions for both the subhalo's light and mass}. However, the dark subhalo model fails to recover the true density profile of the input subhalo, with the concentration being severely overestimated ($\mathrm{log}_{10} c=2.6_{-0.3}^{+0.4}$, versus a true value of $\mathrm{log}_{10} c=1.5$).

In the right three panels of Fig.~\ref{fig: mock_images}, we show the normalized residuals for our three different model fits. Unsurprisingly, in the no-subhalo case, we see correlated residuals in both arcs (caused by both the mass and light of the input subhalo), similar in appearance to the residuals seen when modelling the real dataset. In the right two panels, we can see that those residuals are suppressed by the inclusion of either a dark or luminous subhalo. Based on the fitting statistics, the luminous subhalo model yields a log Bayesian evidence that is 12 higher than the dark subhalo model, indicating a better fit to the data when subhalo light is included, a difference that is not immediately apparent from visual inspection of the residuals.

In summary, our mock test validates our analysis methodology, and suggests that our interpretation of the strong lensing perturber in SDSSJ0946+1006 being a luminous subhalo is consistent with previous work on this system. If the subhalo is luminous, but at a location overlapping with the lensed arc:
\begin{itemize}
    \item The subhalo light is not obviously seen as a distinct component in the lens-light subtracted image;
    \item The residuals are significantly suppressed when including a dark subhalo in the model, but the subhalo's inferred concentration will be anomalously high;
    \item The Bayesian evidence of a model including a luminous subhalo will be higher than that of a model with a dark subhalo.
\end{itemize}
These results highlight the importance of simultaneously modelling the subhalo’s mass and light to avoid biased inferences about its density.



\section{Discussion}\label{sec:discussion}

The results from Sections \ref{sec:results} and \ref{sec:mocks} make a compelling case that the lensing in SDSSJ0946+1006 is perturbed by a luminous subhalo of the main lens, as opposed to the super-concentrated dark subhalo previously reported. As shown in Table~\ref{tab: summary_fittings}, the log Bayesian evidence of the luminous subhalo model is 16 higher than for the dark subhalo model, constituting strong statistical support. Formally, this means that the observed data is over 3 million times more likely to arise in the case that there is a luminous subhalo, than in the case of a dark subhalo. On top of this, given some reasonable priors on the cosmological model and efficiency of star formation, subhaloes of the masses previously reported are expected to be luminous, such that our luminous subhalo model is also a priori more likely.

However, as has been widely discussed in the context of strong lensing subhalo detections \citep[e.g.][]{He2023, Riordan2023}, such formal criteria can overstate the true significance of a result when model misspecification is possible. In particular, simplifying assumptions about the lens galaxy’s mass distribution can produce residual structures that mimic subhalo effects or artificially boost the evidence for one model over another. This is why the community typically demands much higher thresholds \citep[such as a log-evidence increase of 50 in][]{Vegetti2014_SENS} before confidently claiming subhalo detections. We also note that by modelling the second lensed source in this system, \citeMB{} and \citeE{} infer a higher concentration for the dark subhalo. This is likely because the second lensed source provides tighter constraints on the macro model, particularly the multipoles, thereby helping to break some degeneracies between the mass distribution of the main lens and the subhalo. This suggests that, in the luminous subhalo case, there is a potential that a bit higher concentration could still be inferred when modelling the second lensed source, although (e.g. see Figure~6 of \citeE{}) the high-concentration solution tends to be associated with a model that does not include multipoles (as fitted in our case). In light of these considerations, while our result strongly favours the luminous subhalo scenario within our current modelling framework, we caution that it should not be interpreted as irrefutable.


\subsection{Comparing with the luminous subhalo results of \citeMA{}}

We note that there was a previous attempt to jointly fit for the mass and light of a subhalo perturbing the strong lensing in SDSSJ0946+1006, by \citeMA{}. They adopted a Gaussian light distribution for the perturber, and measured the perturber luminosity to be $(1.1\pm0.1) \times 10^8 \,\mathrm{L}_\odot$, with a half-light radius of $0.74^{+0.15}_{-0.10}\,\mathrm{kpc}\,(0.20^{+0.04}_{-0.03}\,\mathrm{arcsec})$. The size and total luminosity of the \citeMA{} perturber light distribution is similar to what we infer ($ 2.5_{-0.5}^{+0.7} \times 10^8 \,\mathrm{L}_\odot$, with a half light radius of $0.77^{+0.30}_{-0.22} \,\mathrm{kpc}$). \citeMA{} also noted that including a subhalo light profile reduced the inferred values of $M_\mathrm{1kpc}$, though they still inferred the subhalo to be compact, with a very small scale radius of $\sim0.01\,\mathrm{kpc}$. 

One possible explanation for the differences between our results and those of \citeMA{} is the flexibility of the respective modelling approaches. In our analysis, the intensities of the MGE components representing the main lens light are optimized while varying the subhalo’s mass and light model. This allows the lens light distribution to adjust in response to potential subhalo emission. In contrast, \citeMA{} pre-subtracted the lens light before fitting the subhalo, which may have caused over-subtraction and, consequently, an underestimation of the subhalo’s light contribution and its associated uncertainties.

\citeMA{} were concerned that the inferred subhalo luminosity of $\sim10^8\,\mathrm{L_\odot}$ could not be trusted. Their worry was that the subhalo light component in their model may have simply removed residuals that had arisen from imperfect lens light subtraction, rather than genuinely captured light coming from the subhalo. However, as discussed above, our modelling optimizes the main lens light, subhalo light and subhalo mass simultaneously. This provides the flexibility to disentangle and account for both components, as well as capture degeneracies between the lens light and subhalo light, or between the subhalo's mass and light. Based on our improved analysis, we think it is likely that the detection of non-zero subhalo luminosity is genuine. 

\subsection{Can subhalo light explain the data at other wavelengths?}

Another concern raised by \citeMA{} about the proposal that the SDSSJ0946+1006 perturber produces significant emission, is the lack of significant emission at the subhalo location in the F336W image. 
This may be explained if the subhalo is, in fact, an old red dwarf satellite galaxy, which emits very little in the ultraviolet. The rest-frame wavelength probed by F336W corresponds to approximately 275 nm, which for old dwarf galaxies is typically 2–6 magnitudes fainter than r-band emission \citep{Kim2010}. Assuming the subhalo is 3 magnitudes fainter in F336W than in F814W, the expected peak pixel flux would be $\sim0.0025\mathrm{e}^-/\mathrm{pix}$. Given the estimated background noise near the subhalo’s location is $\sim0.001\mathrm{e}^-/\mathrm{pix}$, even after a 5700 s exposure (the exposure time of the F336W observation), the peak single-pixel S/N would be merely $\sim2.8$. This would make the signal difficult to observe in the image, especially considering contamination from the main lens light and other sources.

A more robust approach would be to repeat our modelling on the F336W image to place a quantitative constraint on the subhalo's luminosity. In fact, this system has also been observed in three additional HST bands: F438W, F606W, and F160W. If the perturber is indeed a red dwarf satellite, it is likely to exhibit stronger emission at longer wavelengths --- especially in the infrared, such as in F160W. We leave a comprehensive analysis of the subhalo's emission across all available HST bands to future work, but note that such an analysis would likely be conclusive: if the perturber is dark then the same highly-concentrated mass distribution should be inferred at each wavelength (or should at least be consistent with the data at each wavelength, given some wavelengths may be less sensitive to the subhalo's presence and/or properties); if the perturber is luminous, then again there should be a consistent mass distribution inferred across wavelengths, and the perturber luminosity should vary with wavelength in a manner that is reasonable for a low-redshift dwarf galaxy.

\subsection{The subhalo's nature as a dwarf satellite galaxy}
\begin{figure}
    \includegraphics[width=1.0\linewidth]{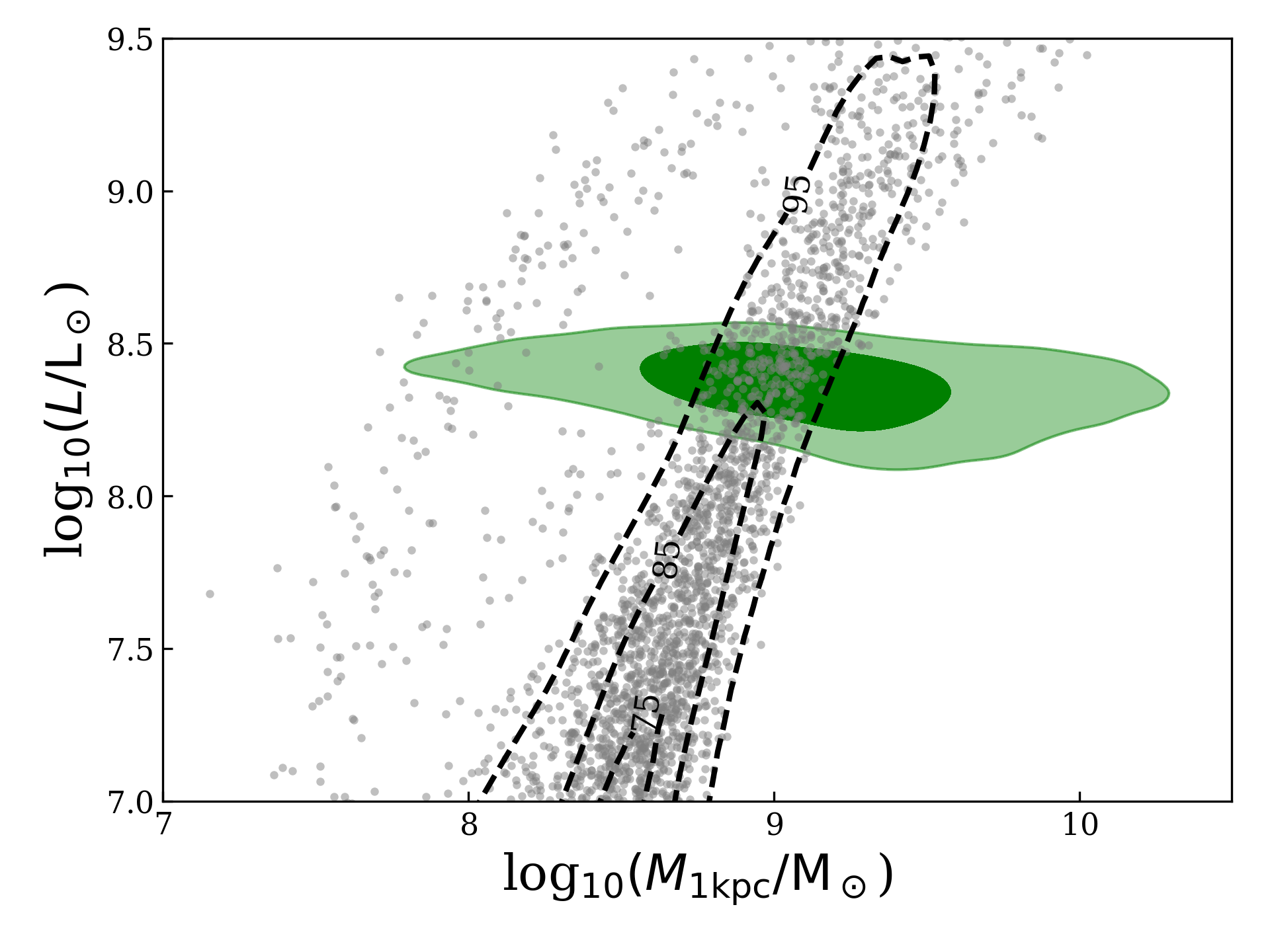}
    \caption{Comparison between our measurment of the SDSSJ0946+1006 subhalo's mass ($M_{\rm 1kpc}$)/luminosity and those of subhaloes in TNG50-1. The green contours indicate our measurement while the gray points represent subhaloes in the simulation with $M_{\rm 1kpc}>10^7{\rm M_\odot}$. The dashed black contours enclose 75, 85 and 95 percent of those simulated subhaloes.}
    \label{fig:lum_mass_comparison}
\end{figure}

In the previous sections, we have shown that after accounting for the possible contribution of a luminous galaxy at the centre of the subhalo, the inferred concentration of the subhalo is consistent with the CDM expectation. The subhalo mass indicates our detection could be a dwarf satellite galaxy in SDSSJ0946+1006. However, to confirm this possibility, an important aspect to examine is whether the inferred subhalo luminosity is reasonable for the detected subhalo mass in the case of a dwarf satellite. This has been thoroughly investigated in \citeD{}, where the authors systematically compared the properties inferred for observed subhaloes with those from the TNG50-1 simulation \citep{Pillepich2018}. In particular, their Figure 9 presents the relationship between the subhalo mass (as defined by \textsc{SUBFIND} \citep{Springel2001}) and V-band luminosity. Similarly, in Fig.~\ref{fig:lum_mass_comparison}, we compare our new observational constraints with subhaloes from TNG50-1. We select out all subhaloes with $M_\mathrm{1kpc}>10^7\,\mathrm{M_\odot}$ from $5\times10^{12}-5\times10^{13}\,\mathrm{M_\odot}$ haloes in TNG50-1 and their associated V-band luminosity. The gray points show their $M_\mathrm{1kpc}-V\,\mathrm{luminostiy}$ and the green contours represent our measurement. As shown, once the light contribution from the subhalo is included, our observational result aligns remarkably well with the CDM prediction. The amount of detected light is entirely reasonable for a subhalo with a mass ($m_{200}$) of $\sim 10^{10}\,\mathrm{M}_\odot$.

\subsection{Degeneracy between the subhalo's mass and light}

Introducing an additional light distribution associated with the subhalo might raise concerns, since subhaloes typically produce modest lensing perturbations. One might worry that the subtle signals produced by the lensing effects from the perturber's mass could be misattributed to the added light component, making it hard to detect the subhalo’s true gravitational effect. However, our mock tests show that this is not an issue in our case. As illustrated in Fig.~\ref{fig: posteriors_mock}, there is no strong degeneracy between the subhalo’s light profile and its mass distribution. The posterior distribution for $m_{200}$ clearly rules out the massless scenario, demonstrating that our method reliably detects both the presence of the light and mass of the subhalo.

Presumably the subhalo's light and mass perturb the image in sufficiently different manners, such that their effects can be  distinguished. However, we note that our test only examines one specific configuration for one lens system. A more systematic exploration would be required in order to understand how this degeneracy behaves under different conditions -- for example, in other regions of the lensing arc, with different quality imaging, or for different combinations of subhalo mass and luminosity. Of course, for subhaloes far from the lensed emission, the perturber light would be distinct from the lensed arc, and we would expect very little perturber-mass/perturber-light degeneracy in the model fits. However, our ability to detect lensing perturbations decreases significantly as the projected subhalo location is moved away from the lensing arcs \citep[e.g.][]{2022MNRAS.510.2480D}, so a sizeable fraction of detectable subhaloes will be at locations where any perturber-light would ``blend into'' the lensed source emission, as we believe is the case in SDSSJ0946+1006.

\subsection{Implication for strong lensing subhalo detection}

Our results from both real data and mock tests show that when there is even faint light associated with a perturbing subhalo, fitting the image using only a dark subhalo can lead to the inference of a highly concentrated subhalo mass distribution. As a result, to accurately recover the subhalo’s properties, the light component must be included in the modelling. However, incorporating both mass and light components into the model significantly increases its complexity -- and therefore computational cost -- which becomes a challenge when searching for subhaloes in large lens samples.

Interestingly, the presence of faint light in subhaloes can actually aid their detectability: fitting the signal with a mass-only subhalo model results in a compact mass clump that stands out clearly and is less degenerate with the mass distribution and surface brightness distribution of the main lens galaxy. Therefore, we suggest that when searching for substructures in large strong lensing samples, it may remain effective to fit dark subhalo models to detect localized perturbations, as long as the mass model allows for variable compactness (e.g., a free NFW concentration). More detailed modelling that includes a subhalo light component can then be applied in follow-up analyses to more accurately measure the subhalo’s mass distribution.

Finally, we emphasize that, according to CDM theory, a low-mass dark matter halo with $m_{200}\lesssim3\times10^8\,\mathrm{M_\odot}$ is not expected to form a galaxy at its centre \citep{Benitez-Llambay2020}. This suggests that when searching for sufficiently small dark matter haloes, one needs not be concerned about ``contamination'' from any galaxy the halo might host.

\section{Conclusions}\label{sec:conclusions}

In this study, we re-analyze the inferred properties of a low-mass perturber in SDSSJ0946+1006, previously reported to be a super-concentrated subhalo inconsistent with Cold Dark Matter (CDM) predictions. Using similar models to previous works -- assuming a dark perturber with an NFW profile -- we reproduce the same super-concentrated solution, with a concentration exceeding CDM expectations by more than $5\mathrm{\sigma}$.

However, when we instead model the perturber as a luminous subhalo -- adding an elliptical S\'ersic light profile to its mass distribution -- we achieve a significantly improved fit to the lensing data. The inferred subhalo properties also shift closer to CDM predictions:
\begin{align*}
\log_{10}\left(m_{200} / \mathrm{M}_\odot \right) &= 9.5^{+0.7}_{-1.1}, \\
\log_{10}c &= 1.7^{+1.2}_{-0.9}, \\
\log_{10}\left(L / \mathrm{L}_\odot\right) &= 8.4^{+0.1}_{-0.2}, \\
\log_{10}\left(M_\mathrm{1kpc}\right) &= 8.9^{+1.1}_{-0.8}
\end{align*}
These values indicate a concentration consistent with CDM and contrast sharply with earlier results that suggested the need for alternative dark matter models, such as self-interacting dark matter with a large interaction cross-section. The detected luminosity is also consistent with that of simulated dwarf satellites hosted by subhaloes of similar mass.

Our results suggest that the perturber could be a dwarf satellite galaxy with a total mass around $10^9$ to $10^{10}\,\mathrm{M}_\odot$. To validate our approach, we generated a mock image resembling the observed F814W data, including a subhalo with similar mass and luminosity as we inferred from the real data. Modelling this mock image with the same analysis pipelines we used on the real data, we find similar results: a dark subhalo model yields an artificially high concentration, while a joint light-and-mass subhalo model leads to accurate recovery of the input properties. This highlights the importance of modelling both the mass and light of subhaloes in lensing analyses.

While the luminous subhalo model yields a statistically better fit and aligns more closely with expectations from galaxy formation theory in a CDM universe, we caution against over-interpreting this result. Lensing analyses are inherently complex and sensitive to model assumptions, and we cannot fully exclude alternative explanations such as a truly dark, highly concentrated subhalo. Further constraints from imaging in other bands will be crucial in definitively establishing the nature of the perturber.

\begin{acknowledgments}
We thank Simona Vegetti, Conor O'Riordan, Maryam Tajalli, Quinn Minor and Wolfgang Enzi for insightful discussions and comments on the manuscript. QH, AA, CSF and SMC acknowledge support from the European Research Council (ERC) Advanced Investigator grant DMIDAS (GA 786910), to C.S.\ Frenk. AR gratefully acknowledges funding from NASA Grant \#80NSSC24M0021, ``Project Infrastructure for the Roman Galaxy Redshift Survey''.

This paper makes use of the DiRAC Data-Centric system, project code dp004 and dp195, which are operated by the Institute for Computational Cosmology at Durham University on behalf of the STFC DiRAC HPC Facility (www.dirac.ac.uk). These were funded by BIS National E-infrastructure capital grant ST/K00042X/1, STFC capital grants ST/H008519/1, ST/K00087X/1, ST/P002307/1, ST/R002425/1, STFC DiRAC Operations grant ST/K003267/1, and Durham University. DiRAC is part of the National E-Infrastructure.
\end{acknowledgments}

\software{
\href{https://github.com/astropy/astropy}{\textsc{Astropy}}
\citep{astropy1, astropy2}, \href{https://bitbucket.org/bdiemer/colossus/src/master/}{\textsc{Colossus}}
\citep{colossus},
\href{https://github.com/dfm/corner.py}{\textsc{Corner.py}}
\citep{corner},
\href{https://github.com/joshspeagle/dynesty}{\textsc{Dynesty}}
\citep{dynesty},
\href{https://github.com/matplotlib/matplotlib}{\textsc{Matplotlib}}
\citep{matplotlib},
\href{numba` https://github.com/numba/numba}{\textsc{Numba}}
\citep{numba},
\href{https://github.com/numpy/numpy}{\textsc{NumPy}}
\citep{numpy},
\href{https://github.com/rhayes777/PyAutoFit}{\textsc{PyAutoFit}}
\citep{Nightingale2021},
\href{https://github.com/Jammy2211/PyAutoGalaxy}{\textsc{PyAutoGalaxy}}
\citep{pyautogalaxy},
\href{https://github.com/Jammy2211/PyAutoLens}{\textsc{PyAutoLens}}
\citep{Nightingale2018, Nightingale2019, Nightingale2021, Nightingale2024},
\href{https://www.python.org/}{\textsc{Python}}
\citep{python},
\href{https://github.com/scikit-image/scikit-image}{\textsc{Scikit-image}}
\citep{scikit-image},
\href{https://github.com/scikit-learn/scikit-learn}{\textsc{Scikit-learn}}
\citep{scikit-learn}.
}



\bibliography{sample7}{}
\bibliographystyle{aasjournalv7}


\appendix

\setcounter{table}{0}   
\setcounter{figure}{0}
\renewcommand{\thetable}{A\arabic{table}}
\renewcommand{\thefigure}{A\arabic{figure}}

\section{Model Priors of the \textit{Subhalo} Phase}

\begin{table}[htp]
    \begin{tabular}{ccccccc} 
    \hline
    \multicolumn{7}{c}{EPL} \\
    \hline
    $x$ & $y$ & $R_{\rm E}$ & $e_{0}$ & $e_{1}$ & $\gamma$ \\
    \hline
    $\mathcal{G}(0.00, 0.05)$ & $\mathcal{G}(-0.02, 0.05)$ & $\mathcal{G}(1.4, 0.4)$ & $\mathcal{N}(0.04, 0.2)$ & $\mathcal{G}(-0.02, 0.2)$ & $\mathcal{G}(2.1, 0.2)$ \\
    \hline
    \multicolumn{7}{c}{External Shear} \\
    \hline
    $\gamma_{1}^\mathrm{ext}$ & $\gamma_{2}^\mathrm{ext}$ \\
    \hline
    $\mathcal{G}(-0.06, 0.05)$ & $\mathcal{G}(0.06, 0.05)$ \\
    \hline
    \multicolumn{7}{c}{NFW Subhalo Mass} \\
    \hline
    $x$ & $y$ & $\log_{10}m_{200}$ & $\log_{10}c$ \\
    \hline
    $\mathcal{U}(-1.63, -0.63)$ & $\mathcal{U}(-0.95, 0.05)$ & $\mathcal{U}(7, 12)$ & $\mathcal{U}(0.2, 3.5)$ \\
    \hline
    \multicolumn{7}{c}{S\'ersic Subhalo Light} \\
    \hline 
    $x$ & $y$ & $q$ & $\phi$ & $\log_{10}I_\mathrm{e}$ & $r_\mathrm{e}$ & $n$ \\
    \hline
    $\mathcal{U}(-1.63, 0.63)$ & $\mathcal{U}(-0.95, 0.05)$ & $\mathcal{U}(0.4, 1.0)$ & $\mathcal{U}(0.0, 180.0)$ & $\mathcal{U}(-3.0, -0.7)$ & $\mathcal{U}(0.05, 0.3)$ & $\mathcal{U}(0.4, 3.0)$ \\
    \hline
    \end{tabular}
    \caption{Priors used for the models used in the \textit{Subhalo} phase. $\mathcal{G}(\mu, \sigma)$ refers to a Gaussian prior centering around $\mu$ with a standard deviation of $\sigma$. $\mathcal{U}(a, b)$ refers to a uniform prior between $a$ and $b$.}\label{tab: priors}
\end{table}

\section{Posteriors of parameters of the luminous subhalo model.}

\begin{figure}[htp]
    \includegraphics[width=1.0\columnwidth]{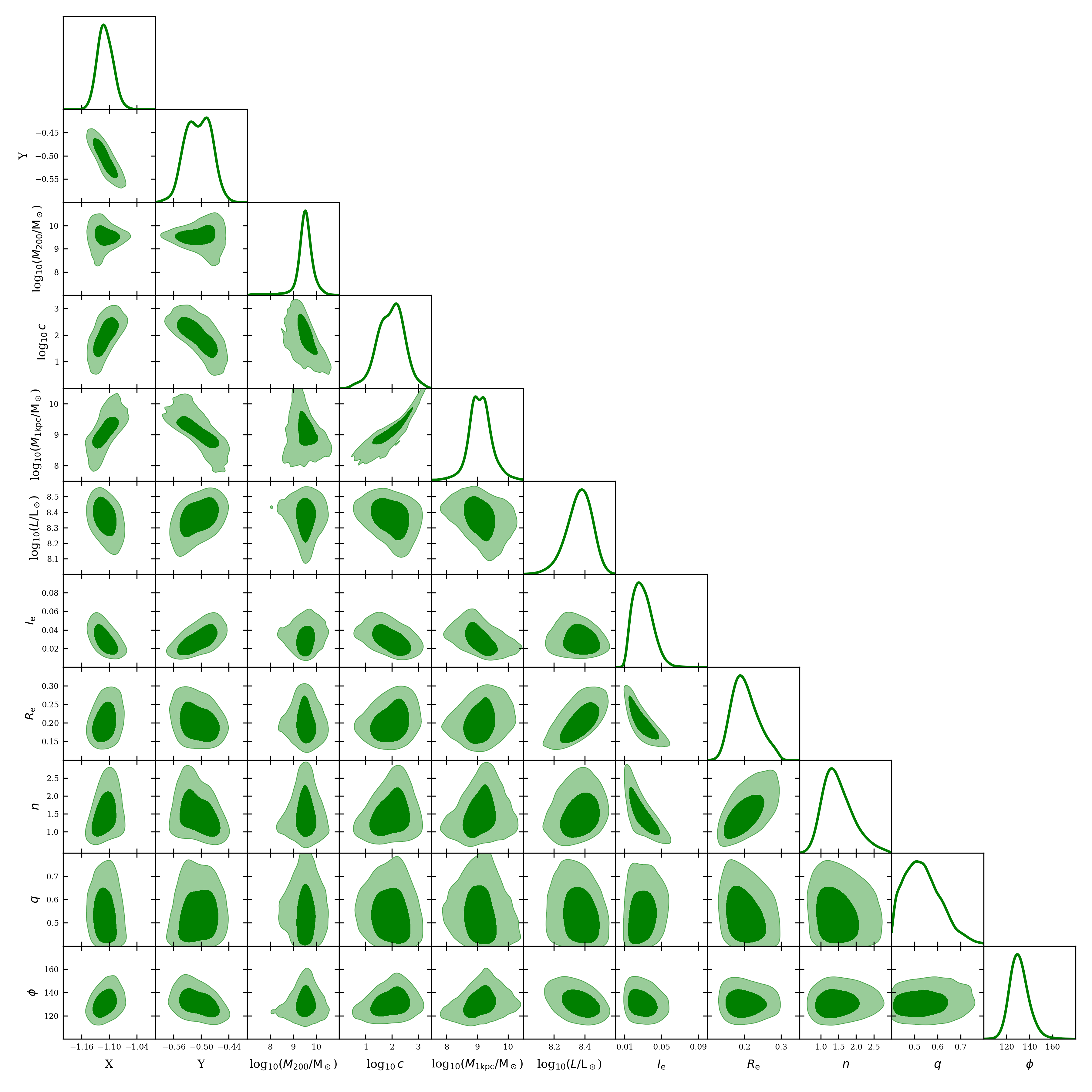}
    \caption{Posteriors of the parameters of the luminous subhalo model. The unit of $X$, $Y$, $R_{\rm e}$ is arcsec, the unif of $I_{\rm e}$ is ${\rm e^-/pix}$ and the unit of $\phi$ is degree. The dark and light green colors indicate the 68\% and 95\% credible intervals respectively.}
    \label{fig: posteriors_subhalo_light}
\end{figure}

\begin{figure}[htp]
    \includegraphics[width=1.0\columnwidth]{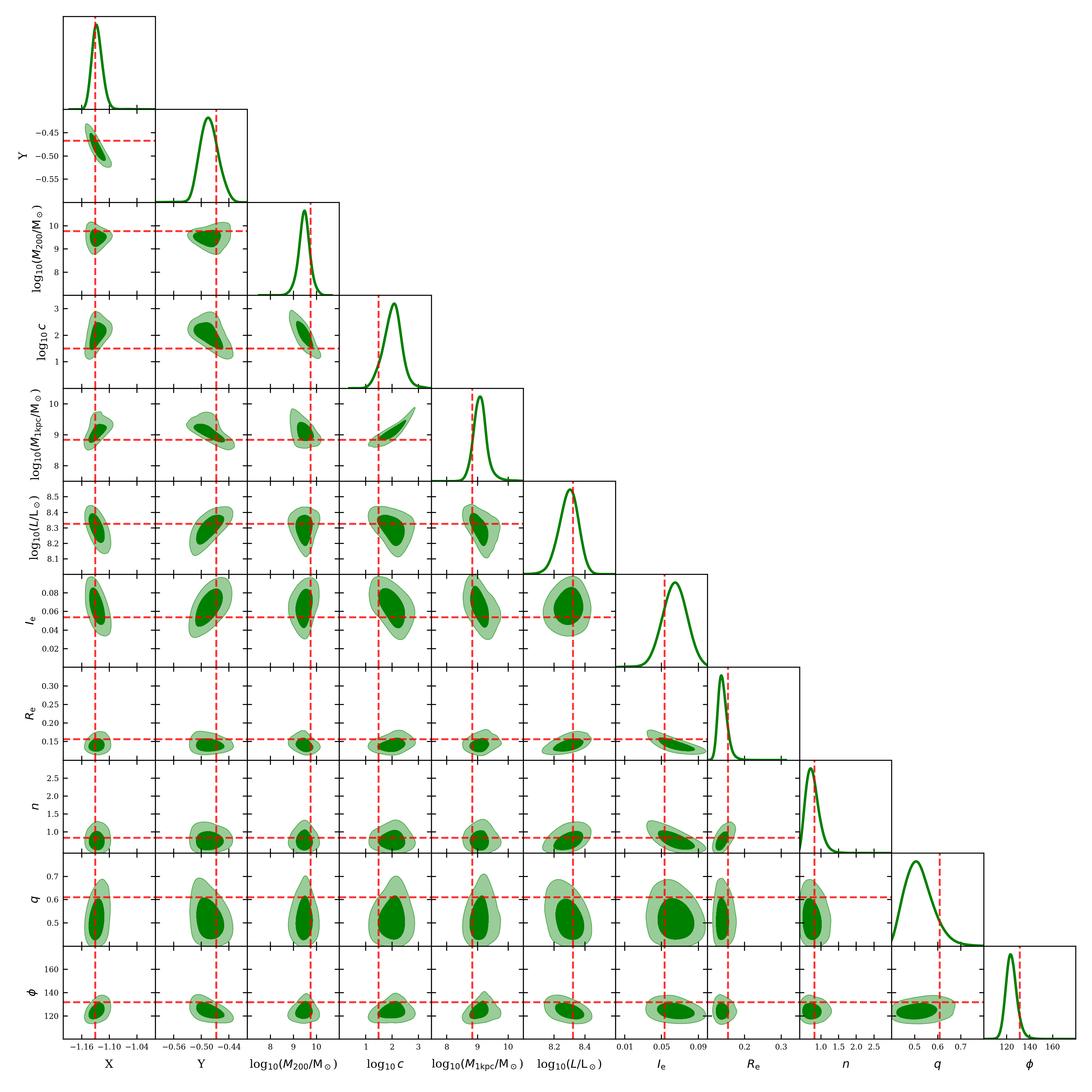}
    \caption{Similar to Fig.~\ref{fig: posteriors_subhalo_light}, but showing the mock test results. The red dashed lines mark the true input values.}
    \label{fig: posteriors_subhalo_light_mock}
\end{figure}

\end{document}